      [1999/12/01 v1.4c Il Nuovo Cimento]
\documentclass{cimento}

\usepackage{graphicx}  
\title{Measuring the Cosmic Microwave Background Radiation (CMBR)
polarization with QUIET}
\author{D.~Samtleben\from{ins:mpifr}\thanks{on behalf of the QUIET collaboration (see \it quiet.uchicago.edu)}}
\instlist{\inst{ins:mpifr} Max-Planck-Institut f\"ur Radioastronomie Bonn
}
\PACSes{
\PACSit{98.80.Es}{Observational cosmology}}
\begin{document}

\maketitle

\begin{abstract}
A major goal of upcoming experiments measuring the Cosmic Microwave
Background Radiation (CMBR) is to reveal the subtle signature of
inflation in the polarization pattern which requires unprecedented
sensitivity and control of systematics. Since the sensitivity of
single receivers has reached fundamental limits future experiments
will take advantage of large receiver arrays in order to significantly
increase the sensitivity. Here we introduce the Q/U Imaging ExperimenT
(QUIET) which will use HEMT-based receivers in chip packages at 90(40)
GHz in the Atacama Desert. Data taking is planned for the beginning of
2008 with prototype arrays of 91(19) receivers, an expansion to 1000
receivers is foreseen. With the two frequencies and a careful choice
of scan regions there is the promise of effectively dealing with
foregrounds and reaching a sensitivity approaching 10$^{-2}$ for the ratio 
of the tensor to scalar perturbations.
\end{abstract}
\section{The status of polarization measurements}
\begin{figure}
\hspace*{1cm} \includegraphics[width=5cm, angle=90]{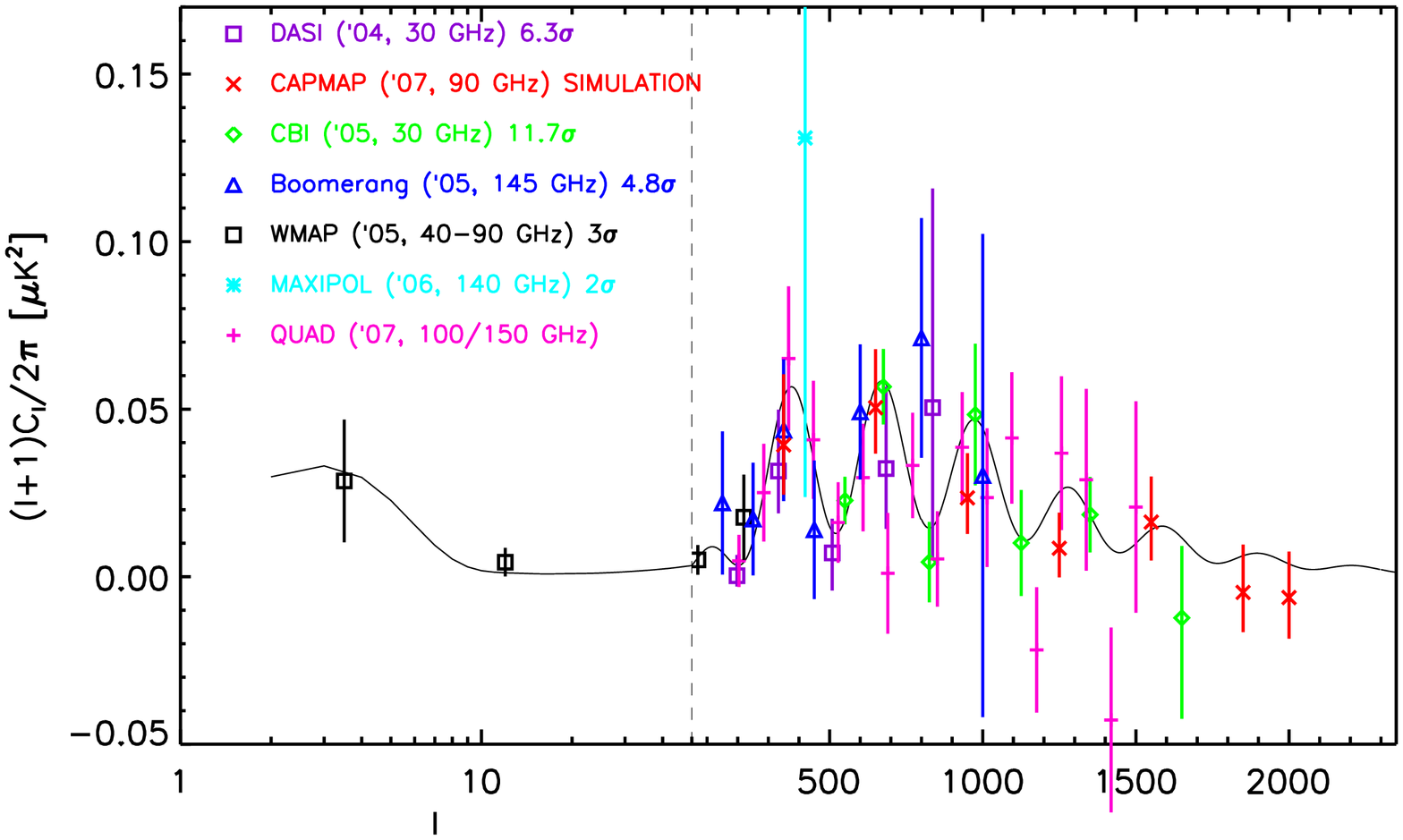}     
\caption{Current measurements of the E-mode power spectrum are shown, the best fit cosmological model is overlaid.}
\label{emodes}
\end{figure}
The intensity anisotropy pattern of the CMBR has already been measured
to an extraordinary precision, which helped significantly to establish
the current cosmological paradigm of a flat Universe with a period of
inflation in its first moments and the existence of the so called
'Dark Energy' \cite{wmapsp}. The polarization anisotropies of the CMBR
are an order of magnitude smaller than the intensity anisotropies and
provide partly complementary information. The polarization pattern is
divided into two distinct components termed E- and B-modes which are
scalar (pseudoscalar) fields. The E-modes originate from the dynamics
due to the density inhomogeneities in the early universe. The B-modes
are caused by lensing of the E-modes by the matter in the line of
sight and by gravitational waves in the inflationary period in the
very early universe and are expected to be at least one order of
magnitude smaller than the E-modes. The status of the E-mode
measurements is summarized in Figure \ref{emodes} from which it
becomes obvious that the measurements are consistent with the
theoretical model but not yet giving meaningful constraints. 
Of special importance and interest are the B-modes
expected from gravitational waves in the inflationary epoch, since a
detection would allow unique access to the very first moments of the
Universe. The size of this contribution cannot be predicted by theory,
but is parametrized by the tensor-to-scalar ratio, $r$
\cite{infl}. Interesting inflationary energy scales of the order of
the Grand Unifying Theory (GUT) scale of 10$^{16}$ GeV correspond to an $r$
of $\sim$10$^{-2}$, which would give rise to detectable signals of a
few 10~nK. The tiny signal requires unprecedented sensitivity and
control of systematics and foregrounds.  By now receivers have reached
sensitivities close to fundamental limits, so that the sensitivity
will only be increased with the number of receivers. 
\section{The QUIET Experiment}
Recent developments at the Jet Propulsion Laboratory (JPL) led to the
successful integration of the relevant components of a
polarization-sensitive pseudo-correlation receiver at 90 and 40~GHz in
a small chip package. This opened the way to future inexpensive mass
production of large coherent receiver arrays and led to the formation
of the Q/U Imaging Experiment (QUIET) collaboration. Experimental
groups from 12 international institutes \footnote{Max-Planck-Institut
f\"ur Radioastronomie Bonn, Caltech, Columbia University, JPL, Kavli
Institute for Cosmological Physics at the University of Chicago, Kavli
Institute for Particle Astrophysics and Cosmology at the Stanford
University, KEK, University of Manchester, University of Miami,
University of Oslo, University of Oxford, Princeton University} have
joined the experiment and are working on the first prototype arrays
which are planned for deployment for 2008 in Chile. A W-band (90 GHz)
array of 91 receivers and a Q-band (40 GHz) array of 19 receivers will
be deployed on new 1.4 m telescopes mounted on the existing platform
of the Cosmic Background Imager (CBI) in the Atacama Desert at an
altitude of 5080~m. It is foreseen to expand the arrays for a second
phase of data taking (2010++) to arrays with 1000 receivers. For the
expansion it is planned to mount more 1.4~m telescopes on the platform
and relocate the 7m Crawford Hill Antenna from New Jersey to Chile to
also access small angular scales.

\begin{figure}
\hspace*{1cm}
\includegraphics[width=4cm]{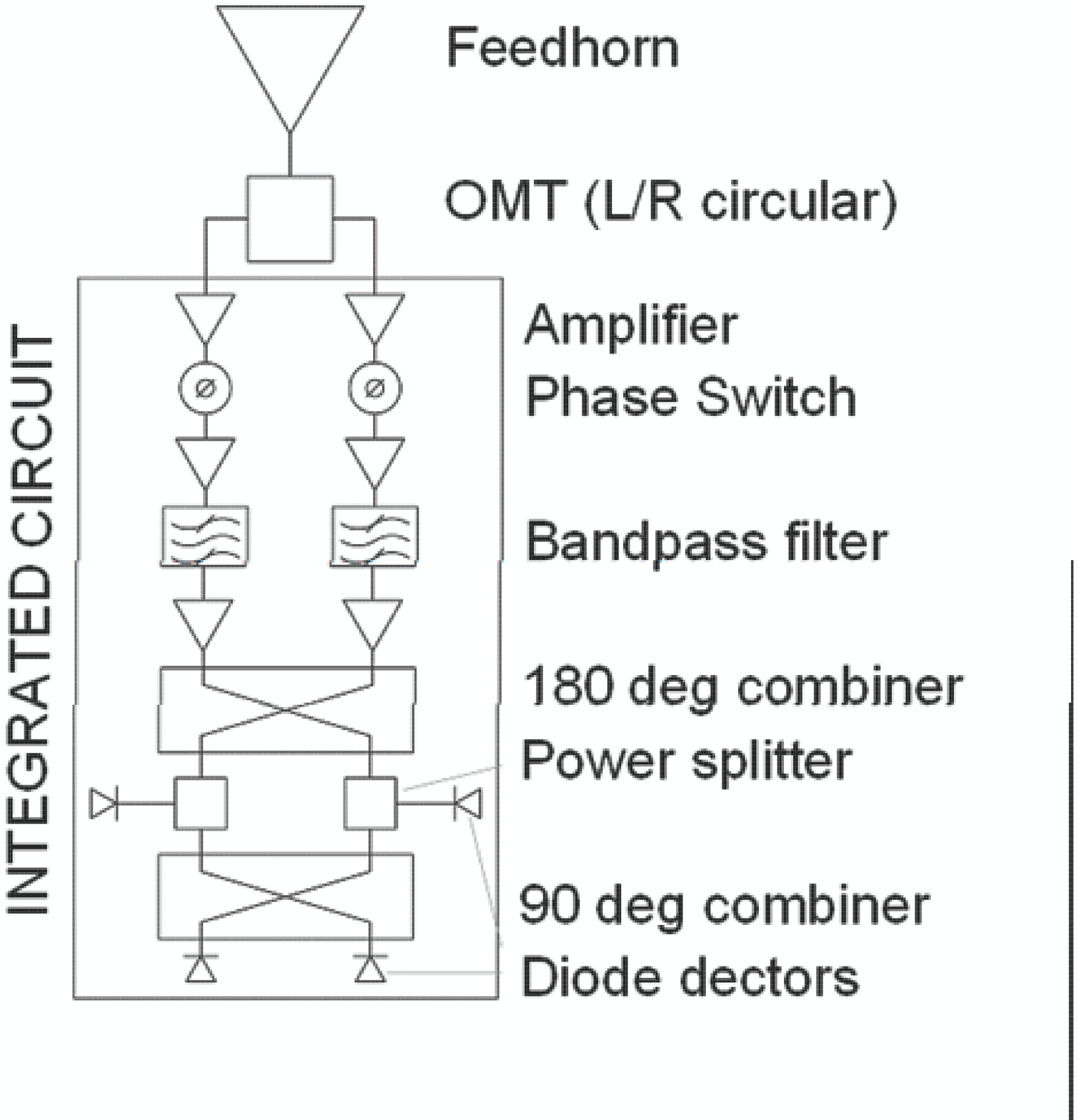}     
\hspace*{1.5cm}
\includegraphics[width=5cm]{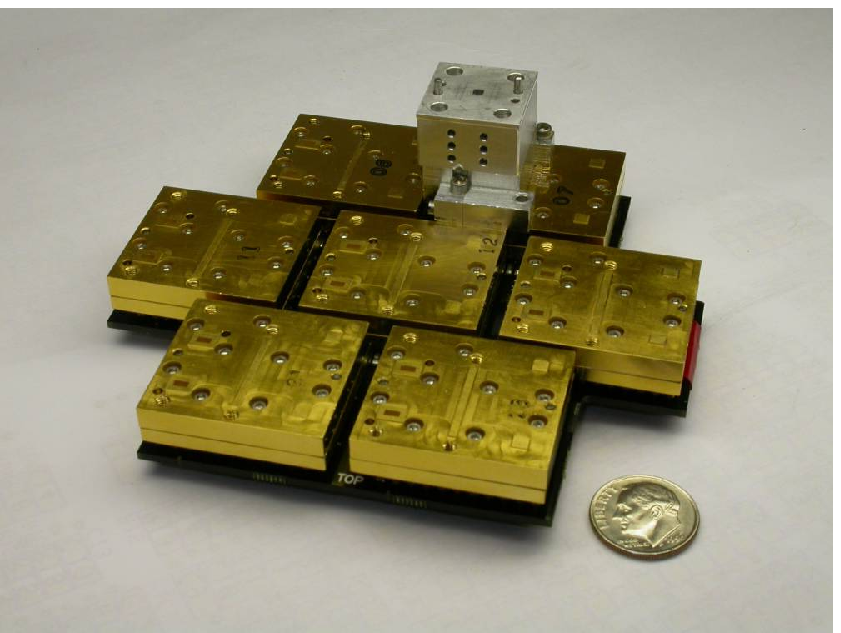}     
\caption{Left: Scheme of a single QUIET receiver. Right: Photo of a 7 element W-band (90 GHz) subarray of the chip receivers with one OMT mounted.}
\label{receiver}
\end{figure}

A sketch of one receiver and its components can be seen in Figure
\ref{receiver}. The incoming radiation couples via a feedhorn to an
Orthomode Transducer (OMT) and from that to the two input waveguides
of the chip package. The chip contains a complete radiometer with High
Electron Mobility Transistors (HEMTs) implemented as Monolithic
Microwave Integrated Circuits (MMICs), phase shifters, hybrid couplers
and diodes. The outputs of the four diodes of the radiometer provide
measurements of the Stokes parameters Q and U and fast (4kHz) phase
switching reduces the effects of the 1/f drifts of the amplifiers. For
10$\%$ of the receivers the OMT will be exchanged by a Magic Tee
assembled in a way that the receivers measure temperature differences
between neighboured feeds. The signals from the diodes are processed
by a digital backend, sampling at 800~kHz with subsequent digital
demodulation. This allows unique monitoring of high-frequency noise as
well as the production of null-data sets with out-of-phase
demodulation giving a valuable check of possible subtle
systematics. The receiver arrays together with the feedhorns are
assembled in large cryostats and the chip radiometers are kept at 20~K
to ensure low noise from the HEMTs.

For a single element a bandwidth of 18(8)~GHz and a noise temperature
of 45 (20) K is aimed for at 90 (40) GHz, leading to expected
sensitivites in Chile of 250 (160) $\mu$K$\sqrt{s}$ per element.

A prototype array of 7 elements with one OMT mounted on top of one
chip radiometer is shown on the right hand side of Figure \ref{receiver}.
The hexagonal prototype arrays of 91 and 19 elements are being
assembled from similar subarrays. The OMTs were built in
cost-effective split-block technique and the corrugated horn arrays
were produced as platelet arrays where 100 plates with feed-hole
patterns are mounted together by diffusion bonding.

The increase in sensitivity is a necessary but not yet sufficient
condition for the successful measurement of B-modes as the signal of
interest is smaller than the one from astrophysical foregrounds. The
diffuse emission (synchrotron, dust) from our galaxy and extragalactic
sources produces polarized signals of which the distribution and
characteristics are not yet known to the precision required for a full
removal. Multifrequency observations are mandatory to study the
foreground behaviour and enable the clean extraction of the CMBR
polarization anisotropies. QUIET in its observations will use two
frequencies which frame the frequency where the contamination from
foregrounds in polarization is expected to be minimal, around
70~GHz. Also, it will coordinate the patches to be observed with other
polarization experiments to gain additional frequency information.
Fields were selected in which minimal foreground contamination is
expected.  The B-modes from gravitational waves will suffer from yet
another foreground (which in intself is of scientific interest)
which is the lensing of E-modes into B-modes. Using
the observations at small angular scales QUIET will be able to
determine a lensing correction and, with that, be able to
remove that contribution properly.
\subsection{Comparison to other Experiments}
\begin{figure}
\includegraphics[width=15cm]{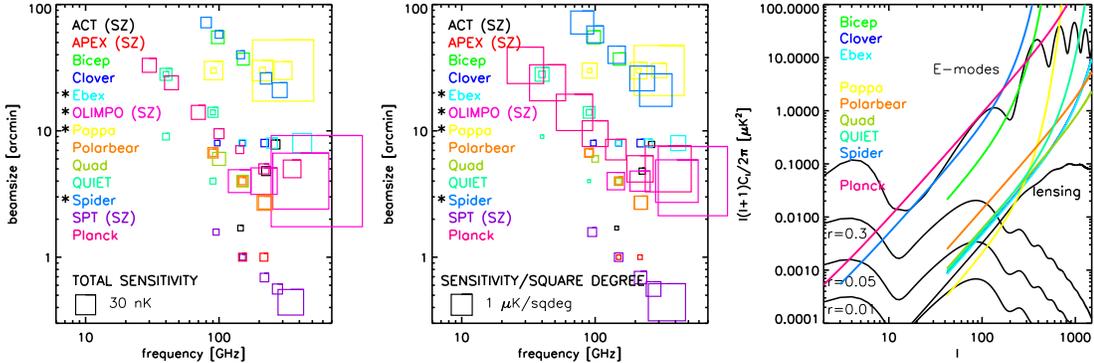}     
\caption{Left: Beamsize versus frequency for the different experiments, the area of the squares corresponds to the total sensitivity. Experiments marked with a '*' are balloon experiments, Planck is the only space mission. Middle: Beamsize versus frequency, the area of the squares corresponds to the sensitivity per square degree. Right: White noise level for the CMBR polarization experiments as function of the multipole l in comparison to the expected E- and B-mode power spectra.}
\label{comparison}
\end{figure}
While currently ongoing CMBR experiments (BICEP, QUAD) are running
with tens of receivers all future experiments are aiming for large
arrays with several hundreds of receivers, all of them (but QUIET)
using bolometers.  Figure \ref{comparison} visualizes the main
parameters of QUIET in comparison to other ongoing and planned CMB
experiments (no interferometers are shown). Some of the experiments
have their main focus on observations of the Sunyaev Zeldovich effect
\footnote{This effect takes advantage of the distortion of the CMBR
spectrum from the scattering of the photons in galaxy clusters to
study cluster physics and the evolution of the Universe} (marked
accordingly) and not polarization observations, but may still upgrade
their detector arrays for polarization sensitivity. The parameters of
the future experiments were taken from recent papers and talks about
the various efforts, but since some of the technologies are not yet
fully established and not all of the experiments are completely
funded, it is clear that some of the parameters may change in
the course of the production. Both the left and middle plot display
beam size versus frequency for the different experiments while the size of
the squares indicates different parameters of the experiments. Since
some experiments (QUIET, Polarbear) are planned to operate in
different phases, they have several squares at the same position. In
the left panel the square area is proportional to the total sensitivity
of the experiments, which means the smaller the square the more
sensitive the experiment. As can be seen the next generation of CMB
experiments will achieve the desired level of a few nK sensitivity.

Except for the space-based mission Planck and the balloon experiment
SPIDER all ground-based experiments focus their sensitivity on small
fractions of the sky. In this way it is possible to avoid regions of
high foreground contamination and also gain a higher signal-to-noise
ratio in the maps, which helps characterizing foregrounds and
systematics. In order to compare the sensitivity on a
map the middle figure displays squares which are in size proportional
to the sensitivity in $\mu$K/square degree. The right figure then
shows the corresponding white noise level for the different
polarization experiments as a function of multipole l in comparison to
the different polarization power spectra. As can be seen the white
noise power is for Planck a factor of 100 higher than for the
ground-based experiments, which means the noise on a QUIET map is
about one order of magnitude lower than on the maps expected from
Planck. The main sensitivity of Planck for the measurement of B-modes
from gravitational waves comes from the reionization peak at low
multipoles of l while the ground-based experiments like QUIET will
constrain $r$ from measuring at the maximum of the gravitational wave
signal at $l$=100, corresponding to an angular scale of 2~degrees.

QUIET is complementary to other experiments in many different ways:
\begin{itemize} 
\item QUIET is the only ground-based effort using coherent receivers and thus
dealing with different systematics than the bolometric systems.
\item It is the only experiment to measure the Stokes parameters Q and U
simultaneously in one pixel which provides a good handle on several
systematic effects.
\item The array at 40 GHz complements the high frequencies of the
bolometer arrays and thus allows to account for the contamination from
synchrotron radiation which dominates at low frequencies.
\item By using different telescope sizes QUIET will be able to measure
both large and small angular scales with the same receivers.
\end{itemize}
\subsection{Science reach}
\begin{figure}
\includegraphics[width=11.9cm]{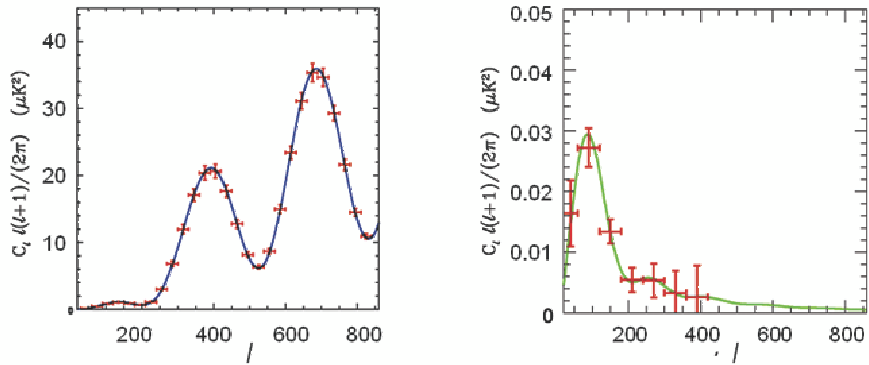}     
\caption{QUIET power spectra as expected for phase II using only the sensitivity of the W-band arrays and the 1.4 m telescopes and including several realistic effects (see text). Left: E-mode spectrum. Right: B-mode spectrum for $r$=0.36 (lensing contribution not shown).}
\label{powspec}
\end{figure}
Already in phase I QUIET will be able to measure the E-mode spectrum
to an unprecedented precision. The expected E- and B-mode power
spectra for the phase II of QUIET with 1000 elements are shown in
Figure \ref{powspec}. Only the sensitivity of the W-band (90 GHz)
arrays from the 1.4~m telescopes was used assuming that the Q-band
sensitivity is used for foreground removal. The results were derived
by including several real-data effects: A realistic observing strategy
has been simulated and the method used in CAPMAP to remove
ground-pickup by mode-removal in single scans has been applied
\cite{capmap1,capmap2}. The simulations also incorporate effects from E-B
leakage where the B-mode measurement is degraded due to the E-mode
signal leakage into the B-mode spectrum due to the finite size of the
observed patch \cite{kendrick}. Additionally, the errors include a 
marginalization over the power in adjacent $l$-bins and for B-modes also 
over E power.

The expected precision on cosmological parameters assuming initial
adiabatic conditions is summarized in table \ref{cosmpar}. Note that
these estimates had been performed before the publication of WMAP
results, but do agree well with the published WMAP parameter errors.
From the table one can see that QUIET will improve the WMAP parameter
errors to a size competitive to the expected precision of Planck.
Adding the QUIET measurements to Planck will only bring a small
improvement in most of the parameters.  However, QUIET will already in
phase I be able to constrain the tensor-to-scalar ratio together with
Planck to a level significantly smaller than Planck is expected
to. Adding QUIET phase II will bring the limit on $r$ down to the
level of 10$^{-2}$.
\begin{table}
  \caption{Expected precision on cosmological parameters in
  percent. For $\Delta r$ the 5$\sigma$ upper limit is presented}
  \label{cosmpar} \begin{narrowtabular}{3cm}{llllll} \hline & A & B &
  C & D & E
\\ \hline
      $\Omega_Bh^2$     & 6 & 4 & 1 & 1 & 1 \\
      $\Omega_Mh^2$     & 8 & 7 & 4 & 2 & 2  \\
      $\Omega_\Lambda$  & 15 & 14 & 8 & 4 & 3\\
      $\tau$     & 34 & 23 & 14 & 7 & 6\\   
      $n_s$     & 4 & 2 & 1 & 1 & 1\\
      $\Delta r$  & 1.35 & 0.021 & 0.009 & 0.042 & 0.009\\
    \hline
\multicolumn{6}{l}{A: WMAP~~~~~~B: WMAP + QUIET Phase I~~~~~~C: WMAP + QUIET Phase I and II} \\
\multicolumn{6}{l}{D: Planck~~~~~~~E: WMAP + QUIET Phase I and II} \\
    \hline
  \end{narrowtabular}
\end{table}
\section{Conclusion}
We are entering an era where probing GUT scale physics is possible.  A
number of experiments are in preparation for seeing the signature of
inflation in the B-modes. Of these QUIET is the only one using
coherent detectors. A convincing discovery of the tiny signal will
need consistent measurements from complementary techniques and
observing frequencies. Already within the next years QUIET will reach
the sensitivity to probe, together with other experiments, interesting
levels of $r$.



\end{document}